\documentclass[12pt]{article}
\usepackage{cite}
\def\bea{\begin{eqnarray}}
\def\eea{\end{eqnarray}}

\def\ba{\beq\new\begin{array}{c}}
\def\ea{\end{array}\eeq}

\newcommand{\beq}{\begin{equation}}
\newcommand{\eeq}{\end{equation}}
\newcommand{\beqa}{\begin{eqnarray}}
\newcommand{\eeqa}{\end{eqnarray}}

\newcommand{\Tr}{{\rm Tr}\,}

\newcommand{\ntwo}{\mbox{${\cal N}\!\!=\!2\;$}}
\newcommand{\none}{\mbox{${\cal N}\!\!=\!1\;$}}

\begin{document}
\begin{titlepage}
\renewcommand{\thefootnote}{\fnsymbol{footnote}}

\begin{flushright}
TPI-MINN-00/13\\
UMN-TH-1847/00\\
ITEP-TH-16/00\\
hep-th/0004087\\

\end{flushright}

\vfil

\begin{center}
\baselineskip20pt
{\bf \Large Deconfinement at the Argyres-Douglas point in\\ SU(2)  gauge 
theory with  broken \mathversion{bold}\ntwo  supersymmetry }
\end{center}

\vfil

\begin{center}
{\large A. Gorsky$^a$, A. Vainshtein$^b$, and A. Yung$^c$}

\vspace{0.3cm}

$^a${\it Institute of Experimental and Theoretical Physics, Moscow
117259}\\[0.2cm]
$^b${\it Theoretical Physics Institute, University of Minnesota,
Minneapolis, MN 55455\\[0.2cm]}
$^c${\it Petersburg Nuclear Physics Institute, Gatchina, St. Petrsburg 188350}

\vfil

{\large\bf Abstract} \vspace*{.25cm}
\end{center}

We consider chiral condensates in  SU(2) gauge theory 
with broken \ntwo super\-symmetry. The matter sector contains 
an adjoint multiplet
and one   fundamental flavor. Matter and gaugino condensates 
are determined by integrating out the adjoint field. The only
nonperturbative  input  is the Affleck-Dine-Seiberg (ADS)  superpotential
generated by one instanton plus the Konishi anomaly. These results are
consistent with those obtained by the `integrating in' procedure,
including a reproduction of the Seiberg-Witten curve from the ADS 
superpotential. We then  calculate monopole, dyon, and charge
condensates using the Seiberg-Witten approach. We  show that the
monopole and charge condensates  vanish at  the Argyres-Douglas point where
the monopole and charge vacua collide. We interpret this phenomenon
as  a deconfinement of electric and magnetic charges at  the
Argyres-Douglas  point.

\vfil
\begin{flushleft}
April 2000
\end{flushleft}
\end{titlepage}

\section{Introduction}

The derivation of exact results in \none supersymmetric gauge theories
based  on low energy effective superpotentials and holomorphy was pioneered
in~\cite{ads,SV}, then new wave of development was initiated by
Seiberg,
see~\cite{IS} for review.  
Additional  input was provided by the Seiberg-Witten solution 
 of \ntwo supersymmetric gauge theories with and without
matter~\cite{sw}.

The key feature of the \ntwo theory  is the existence of the Coulomb
branch where the  vacuum expectation value of the adjoint scalar serves as a
modulus \cite{sw}. The solution is described in terms of  Riemann 
surfaces and the Coulomb branch parametrizes the  moduli space of
their  complex structures. 
The simplest way to break  \ntwo supersymmetry (SUSY)
 down to \none amounts to
giving  a nonvanishing   mass $\mu$ to the chiral \none su\-per\-field
in  the adjoint representation. This field is a
partner  to the gauge fields in the \ntwo super\-multiplet. At small
values of  $\mu$ the theory is close to its \ntwo counterpart while at
large  $\mu$ the adjoint matter decouples and the pure \none theory
emerges.  The emerging  theory at large $\mu$ is close to 
supersymmetric  QCD (SQCD) but  does not coincide with 
it. A trace of the massive adjoint remains in the effective theory in the
 form of
nonrenormalizable quartic terms~\cite{kutasov} in the superpotential
which are  suppressed by $1/\mu$.
Although in the \none theory the degeneracy on the Coulomb branch is lifted by
the superpotential, 
 memory of the structure of the Riemann surfaces remains.
Namely, the vanishing of the 
discriminant of the  Riemann surface defines the set of vacua in
the corresponding \none theory~\cite{sw,kutasov,is,giveon,kitao,kt}. 

Different vacua are distinguished by the values of chiral condensates, such
as  the gluino condensate $\langle\Tr\lambda \lambda\rangle$ and the
condensate  of  fundamental matter $\langle\tilde{Q} Q\rangle$.
Generically, the latter 
 can be found in SQCD using the effective superpotential, while the
gluino condensate can be evaluated using the Konishi anomaly 
which relates the  two
condensates (see Ref.~\cite{sv} for a review). To obtain
 the condensate in pure \none
Yang-Mills theory one has to start with the massive SQCD and use the
holomorphy to decouple  massive matter. Recently, some points
concerning   formation
of the condensate and  identification of the relevant field
configurations
were clarified in \cite{hol,davies,rv,konishi}

The brane picture provides another approach to the problem. The  brane
configurations for  \none theories with different matter content are known
\cite{N_one} and the recipe for calculating  minima of the superpotentials
has been formulated \cite{super}. The key point concerning the brane
configurations is that to break SUSY down to \none one has to rotate the \ntwo
picture.  However only configurations which correspond to the
vanishing of the  discriminant
 can be rotated which means that at any value of the adjoint mass these
points remain intact.
The superpotentials calculated from 
brane configurations are in correspondence with  field theory
expectations.

In this paper we consider an \none theory with both adjoint and fundamental
matter and limit ourselves to the most
 tractable case of SU(2) gauge group with
one fundamental flavor and one multiplet in the adjoint representation.
Our strategy is as follows: First, we integrate out the adjoint matter to get 
 SQCD-like effective superpotential for the fundamental matter. The only
nonperturbative input in this effective superpotential is given by the   
Affleck-Dine-Seiberg superpotential generated by one instanton 
\cite{ads}.  Difference 
with  pure SQCD is due to the tree level nonrenormalizable term generated by
the heavy adjoint exchange, mentioned above.  Similarly to SQCD, the
effective  superpotential 
together
with the Konishi relations unambiguously fixes condensates of fundamental and 
adjoint
matter as well as the gaugino condensates in all three vacua of the theory.

We then compare  the condensate of the adjoint matter with 
points in the $u$ plane corresponding to the  vanishing of the discriminant 
defined by Seiberg-Witten solution in \ntwo theory and find a complete
match.
Our results for matter and gaugino condensates are consistent with those 
obtained
by the `integrating in' method~\cite{ils,Intrilligator,giveon} and can be
viewed  as an
independent confirmation of this method. 
What is specific to our approach is that
we start from the  weak coupling regime where the notion of an effective
Lagrangian is  well
defined, and  then use holomorphy to extend results 
for chiral condensates into
strong coupling. 

 We subsequently determine monopole, dyon, and charge 
condensates following  the Seiberg-Witten approach, i.e. considering 
effective superpotentials near singularities on the Coulomb branch of the 
 \ntwo theory. Again, holomorphy allows us to extend our results to the
domain of the ``hard'' \ntwo breaking. This extension include not only
the mass term of adjoint but also  breaking of \ntwo in Yukawa couplings.

Our next step is the study  chiral condensates  in the Argyres-Douglas (AD)
points. These points were originally introduced in the moduli/parameter
space of \ntwo theories as 
points  where two singularities on the Coulomb branch
coalesce~\cite{ad,apsw,hori}. It is believed that the theory in
 the AD point flows in the infrared to a nontrivial
superconformal theory. The notion of the AD point continues to
make sense even when the  \ntwo theory  is broken to \none; in
the \none theory it is the point in parameter space where 
 two vacua collide.

In particular, we consider the AD point where the monopole and charge
vacua collide at a particular
 value of the mass of the fundamental flavor. Our key result is
that  both monopole and charge condensates  vanish at the
 AD  point~\footnote{Vanishing of condensates for coalescing vacua was
   mentioned by Douglas and Shenker~\cite{DS} in the context of SU($N$)
  theories without fundamental matter for $N\ge 3$. Note, that it
  was before the notion of the AD point was introduced~\cite{ad}. }.  
We interpret this  as deconfinement of both
electric and magnetic charges at the  AD
point. It provides evidence that the theory
at the AD point remains superconformal even after strong breaking of
\ntwo to \none. Argyres and Douglas conjectured this in their
consideration of SU(3) theory~\cite{ad} 

Let us recall that the condensation of monopoles ensures confinement
of quarks in the monopole vacuum~\cite{sw}, while the condensation of
charges provides confinement of monopoles in the charge vacuum.
As  shown by 't~Hooft~\cite{H} it is impossible for these two
phenomena  to coexist. This apparently leads to a paradoxical 
situation in the AD
 point where the monopole and charge vacua collide.
Our result resolves this paradox.

This paradox is a part of more general problem: whether there is an uniquely 
defined theory at the AD point. Indeed, when two vacua collide the
Witten index of the emerging effective theory at the AD point is fixed,
namely there are two bosonic
vacuum states. The question is whether there is any physical quantity which
could serve as an order parameter differentiating these two vacua.
The continuity of chiral condensates in the AD point we find 
shows that these condensates are not playing this role. The same continuity 
also leads  to vanishing  tension for domain walls interpolating
between colliding vacua when we approach the AD point. We discuss if
these domain walls could serve as a signal  of two vacua in the AD point.

The paper is organized as follows. In Sec.~\ref{sec:matgaug} we dwell on 
the calculation
of matter and gaugino condensates, while  monopole, charge and dyon 
condensates are
considered in Sec.~\ref{sec:mcd}. 
In Sec.~\ref{sec:walls} we briefly discuss a definition of the theory 
at the AD point and the related problem of domain walls. 
Our results are discussed in Sec.~\ref{sec:disc}.

\section{Matter and gaugino  condensates }\label{sec:matgaug}
\subsection{Effective superpotential and condensates}
\label{sec:sup}

We consider a \none theory with SU(2) gauge group where the matter
sector consists of the adjoint field $\Phi^\alpha_\beta=\Phi^a
(\tau^a/2)^\alpha_\beta$ ($\alpha,\beta=1,2;~a=1,2,3$), and two fundamental 
fields
$Q^\alpha_f$ $(f=1,2)$ describing one flavor.
The  general renormalizable superpotential for this theory has the form,
\begin{equation}
{\cal{W}}= \mu \,{\rm Tr} \,\Phi^2 + \frac{m}{2} \,Q_{ f}^\alpha Q^{f}_\alpha  
+\frac{1}{\sqrt{2}}\,h^{fg} \,Q_{\alpha f}\Phi^\alpha_\beta Q^\beta_g\;.
\label{superp}
\end{equation}
Here the parameters $\mu$ and $m$ are related to the masses of the adjoint and
fundamental fields, 
$m_\Phi=\mu/Z_\Phi$, $m_Q=m/Z_Q$, by the corresponding $Z$ factors in 
the kinetic
terms. Having  in mind normalization appropriate for
 the \ntwo case we choose for bare parameters
$Z_\Phi^0=1/g_0^2$, $Z_Q^0=1$. The matrix of Yukawa couplings
 $h^{fg}$ is  symmetric, and  summation over color indices $\alpha,\beta=1,2$
is explicit. Unbroken \ntwo SUSY appears when $\mu=0$ and $\det h=-1$.

To obtain an effective theory similar to SQCD
 we integrate out the adjoint field
$\Phi$ implying  that $m_\Phi\gg m_Q$. In the classical approximation this 
integration 
reduces to  the substitution
\begin{equation}
\Phi^\alpha_\beta= -\frac{1}{2\sqrt{2}\,\mu}
\,h^{fg}\left( Q_{\beta f}Q^\alpha_g
-\frac 1 2\, \delta^\alpha_\beta \, Q_{\gamma f}Q^\gamma_g\right)\,,
\label{substitute}
\end{equation}
which follows from $\partial {\cal{W}}/\partial \Phi =0$.  What is the effect
 of
quantum corrections on the effective superpotential? It is well known from the
study of SQCD that perturbative loops do not contribute and nonperturbative 
effects
are exhausted  by the  Affleck-Dine-Seiberg (ADS)
superpotential generated by one instanton~\cite{ads}. 
The
effective superpotential then is
\begin{equation}
{\cal{W}}_{\rm eff}= m\, V - \frac{(-\det h)}{4\mu}\, V^2 +\frac{\mu^2
\Lambda_1^3}{4\,V}
\label{sup1}
\end{equation}
where the gauge and subflavor invariant chiral field $V$ is defined as
\begin{equation}
V=\frac 1 2 \,Q_{ f}^\alpha Q^{f}_\alpha \;.
\end{equation}
The first two terms in  Eq.~(\ref{sup1}) appear
 on the tree level after substitution~(\ref{substitute}) into 
Eq.~(\ref{superp})  while the third nonperturbative one is the  ADS
superpotential.  The scale parameter
$\Lambda_1$ is given in terms of the mass of 
Pauli-Villars regulator $M_{\rm PV}$ and
the bare coupling $g_0$ (plus the vacuum angle $\theta_0$) as
\begin{equation}
\Lambda_1^3=4\, M_{\rm PV}^3\exp\left(-\frac{8\pi^2}{g_0^2}+i\theta_0\right)\;.
\end{equation}
The coefficient $\mu^2 \Lambda_1^3/4$ in the ADS superpotential is 
equivalent to $\Lambda^5_{\rm SQCD}$ in SQCD.  The factor $\mu^2$ in the
coefficient reflects four fermionic  zero modes of 
the adjoint field, see e.g. 
Ref.~\cite{y,rv} for details.

The only term in the superpotential~(\ref{sup1}) which differentiates it from
the SQCD case  is  the second term  which is due to tree level
exchange by the adjoint field. At 
$h=0$ it vanishes and we are back to the known SQCD case with two vacua and 
a Higgs phase for small $m$. 

When $\det h$ is nonvanishing we have three vacua, marked by the vevs of 
the lowest component of $V$,
\begin{equation}
v=\langle \,V \,\rangle\,.
\end{equation}
These vevs are roots of 
 the algebraic equation ${\rm d}{\cal W}_{\rm eff}/{\rm d}v=0$ which 
has the form
\begin{equation}
m-\frac{(-\det h)}{2}  \, \frac{v}{\mu}- \frac{\Lambda_1^3}{4} \left(
\frac{\mu}{v}\right)^2=0\;.
\label{vaceq}
\end{equation}
This equation shows, in particular, that although the second term in the
superpotential (\ref{sup1}) seems to be suppressed at large $\mu$ 
it turns out to be   of the same
order as the ADS term. From Eq.~(\ref{vaceq}) it is also clear that the 
dependence
on
$\mu$ is  given by the scaling $v\propto\mu$. 

To see the dependence on the other
parameters let us substitute $v$ by the dimensionless variable $\kappa$
defined by the relation 
\begin{equation}
v=\mu\,\sqrt{\frac{\Lambda_1^3}{4m}}\,\kappa\;.
\label{vk}
\end{equation}
Then Eq.~(\ref{vaceq}), when rewritten in terms of $\kappa$,
\begin{equation}
1-\sigma\, \kappa -\frac{1}{\kappa^2}=0
\label{kappa}
\end{equation}
is governed by the dimensionless parameter $\sigma$,
\begin{equation}
\sigma= \frac{(-\det h)}{4} \,\left(\frac{\Lambda_1}{m}\right)^{3/2}.
\label{sigma}
\end{equation}
We see that the two parameters $m$ and $\det h$ enter only as $m\,(-\det
h)^{-2/3}$. The dependence of $v$ on $\mu$ is linear as we discussed above. 

The particular dependence of condensate $v$ on the parameters
 $\mu$, $m$ and $\det h$  follows from
the  $R$ symmetries of the theory. Following
Seiberg~\cite{seibR} one can consider $\mu$,  $m$ and $\det h$ 
 as background  fields and identify  nonanomalous
$R$ symmetries which prove the dependence discussed above.
Classically, there are three U(1) symmetries in the theory associated with 
the three fermion fields (gaugino, adjoint and fundamental fermions). 
In the quantum theory one can organize two 
nonanomalous combinations  (a symmetry is nonanomalous
if it does not transform the scale $\Lambda_1$, associated with regulators).

 The charges
of the fields and parameters of the theory under these two U(1) symmetries
are shown in Table 1.
\begin{table}[h]
\begin{center}
\begin{tabular}{|c|c|c|c|c|c|c|c|}
\hline
~ & ~  & ~  &~ & ~ & ~ & ~ & ~ \\[-0.1cm]
{\rm Fields/parameters} &  $\Phi$ & $ Q$ &  $W$ &$\theta$ &$ m$ & $\mu$ & $h$ 
 \\[0.2cm]
\hline
\vspace*{-0.2cm}
~ & ~  & ~  &~ & ~ & ~ & ~ & ~\\
U$_{J}(1)~{\rm charges}$ & 0 & $1$ & $1$   & $1$ & 0   & $2$   
 &0\\[0.2cm]
\hline
\vspace*{-0.2cm}
~ & ~  & ~  &~ & ~ & ~ & ~ & ~\\
U$_{R}(1)~{\rm charges} $ & 1 & $-1$ & 1   & 1 & 4   & 0       &3\\[0.2cm]
\hline
\end{tabular}
\caption{Nonanomalous U(1) symmetries}
\label{tabU}
\end{center}
\end{table}
The first of these symmetries U$_{J}(1)$ is a subgroup
of the global SU$_{R}(2)$ group related to the ${\cal N}=2$ 
superalgebra~\cite{sw}. This explains the zero charge of the coupling  $h$ 
 with respect to this symmetry. 
The symmetry U$_{J}(1)$ fixes the $\mu$  dependence of condensates.
Namely,  it is given by a power of $\mu$
equal to  half  the U$_{J}(1)$ charge of the condensate.
 In particular, the field $V$ has   U$_{J}(1)$ charge equal to 2
 which ensures that $v\propto \mu$. Thus,
 we can use holomorphy to extend results to arbitrary values of $\mu$.

The second nonanomalous symmetry U$_{R}(1)$ is similar to the $R$ symmetry
of Ref.~\cite{ads} extended to include the adjoint field. 
As a consequence, for a given chiral field $X$
\begin{equation}
  \label{mdep}
  \langle X \rangle = \mu^{Q_J/2} m^{Q_R/4}\Lambda_1^{d_X
    -(Q_J/2)-(Q_R/4)} f_X(\sigma)\;, 
\end{equation}
where ${Q_J}$,  ${Q_R}$ are the U$_J (1)$, U$_R (1)$ charges of the
field $X$, $d_X$ is its dimension, and $f_X$ is an arbitrary function 
of the dimensionless parameter $\sigma$ defined by Eq.~(\ref{sigma}).
This parameter is neutral under both U(1)'s. The equation~(\ref{vk})
is an example of the  general relation~(\ref{mdep}) with $Q_J=2$, $Q_R=-2$
and $f_V=\kappa(\sigma )/2$.

The important benefit of the consideration above is that in a theory 
with \ntwo SUSY strongly broken by large $\mu$
and $\det h \neq -1$ we  can still relate chiral condensates with those
in softly broken 
\ntwo where $\det h=-1$ and  $\mu$ is small.

Here is an example.
When $\sigma\to 0$ two roots of Eq.\ (\ref{kappa}) are $\kappa_{1,2}=\pm1$ and 
the third one goes to infinity as $\kappa_{3}= 1/\sigma$.
For two finite roots one can suggest dual interpretations.
 Firstly, taking $h=0$,
one can relate them to two vacua of SQCD in the Higgs phase. Second, for $\det
h=-1$ (which is its \ntwo value) one can make  $\sigma$ small by taking the
limit of large $m$. But this limit should bring us  to the monopole and dyon
vacua of softly broken \ntwo SYM. The naming of vacua refers to the particle 
whose
mass vanishes in the corresponding vacuum.

To verify this interesting mapping we need to determine the  vev  
\begin{equation}
u=\langle U\rangle=\langle{\rm Tr} \,\Phi^2\rangle\;,
\end{equation}
which can be accomplished using the set of Konishi anomalies. 
Generic equation for an arbitrary matter field $Q$
 looks as follows (we are using
the notation of the review~\cite{sv}):
\begin{equation}
\frac{1}{4}\,\bar{D}^{2}J_Q= Q\,\frac{\partial \,{\cal{W}}}{\partial
Q}+T(R)\,\frac{{\rm Tr}\,W^2}{8\pi^2}\;,
\end{equation}
where $T(R)$ is the Casimir in the matter representation. The left
hand side is a total derivative in superspace so its average over 
any supersymmetric
vacuum vanishes. In our case this
 results in two relations for the  condensates,
\begin{eqnarray}
\left\langle \frac{m}{2}\,Q_{ f}^\alpha Q^{f}_\alpha +
\frac{1}{\sqrt{2}}\,h^{fg}
\,Q_{\alpha f}\Phi^\alpha_\beta Q^\beta_g+
\frac12 \,\frac{{\rm Tr}\,W^2}{8\pi^2}
\right\rangle=0 \nonumber\\
\left\langle 2\,\mu {\rm Tr} \,\Phi^2 + \frac{1}{\sqrt{2}}\,h^{fg} \,Q_{\alpha
f}\Phi^\alpha_\beta Q^\beta_g +2\,\frac{{\rm Tr}\,W^2}{8\pi^2} 
\right\rangle=0
\label{konishirel}
\end{eqnarray}
From the first relation, after the 
substitution in (\ref{substitute}) and comparing 
with
Eq.\ (\ref{vaceq}), we find an expression for gluino condensate~\cite{NSVZ}
\begin{equation}
s=\frac{\langle {\rm Tr}\,\lambda^2\rangle}{16\pi^2}=-\frac{\langle {\rm
Tr}\,W^2\rangle}{16\pi^2}=\frac{\mu^2 \Lambda_1^3}{4\,v}\;.
\end{equation}
This is consistent with the general expression $[T_G -\sum T(R)]\langle {\rm
Tr}\lambda^2\rangle/16\pi^2$ for the nonperturbative ADS piece of the
superpotential (\ref{sup1}), see~\cite{ds}. Combining the two relations in
(\ref{konishirel}) we can express  the condensate  $u$ in terms of $v$, 
\begin{equation}
u=\frac{1}{2\mu}\left(m\,v +3\,s\right)=
\frac{1}{2\mu}\left(m\,v +\frac 3 4\,\frac{\mu^2 \Lambda_1^3}{v}
\right)=
\frac{\sqrt{m\Lambda_1^3}}{4}\left(\kappa+\frac{3}{\kappa}\right)\;.
\label{ukappa}
\end{equation}
Now we see that in the limit of large $m$ two vacua $\kappa=\pm 1$ are in 
perfect
correspondence with $u=\pm \,\Lambda_0^2$ for the monopole and dyon vacua of
\ntwo SYM. Indeed,
$\Lambda_0^4=m\Lambda_1^3$ is the  correct relation between the 
scale parameters of the theories.

For the third vacuum at large $m$ the value $u=m^2/(- \det h)$
corresponds on the Coulomb branch to the so called charge vacuum, where
some fundamental fields become massless.  Moreover,  the correspondence
with \ntwo results can be demonstrated for the 
three vacua at any value of $m$. To
this end  we use the relation 
(\ref{ukappa}) and Eq.~(\ref{kappa}) to derive the
following equation for
$u$,
\begin{equation}
(-\det h)\,u^3-m^2\,u^2-\frac 9 8 \,(-\det
h)\,m\Lambda_1^3\,u+m^3\Lambda_1^3 +\frac{27}{2^8}\,(-\det
h)^2\Lambda_1^3=0\;.
\label{uone}
\end{equation}
The three roots of this equation are the vevs
 of $\Tr \,\Phi^2$ in the corresponding  vacua.

How does this look from \ntwo side?  The Riemann surface
governing the Seiberg-Witten solution is given by the curve \cite{sw}
\begin{equation}
y^2=x^3-u\,x^2+ \frac{1}{4}\Lambda_{1}^3 m\,x-\frac{1}{64}\Lambda_{1}^{6}\;.
\label{swu}
\end{equation}
Singularities of the metric, i.e. the points in the u-plane where the
discriminant  of the curve vanishes are
defined by
two equations, $y^2=0$ and ${\rm d}y^2/{\rm d}x=0$,
\begin{equation}
x^3-u\,x^2+ \frac{1}{4}\Lambda_{1}^3 m\,x-\frac{1}{64}\Lambda_{1}^{6}=0\;,
\quad 3x^2 -2u\,x + \frac{1}{4}\Lambda_{1}^3 m=0\;,
\end{equation}
which lead to
\begin{equation}
u^3-m^2\,u^2-\frac 9 8 \,m\Lambda_1^3\,u+m^3\Lambda_1^3
+\frac{27}{2^8}\,\Lambda_1^3=0\;.
\label{utwo}
\end{equation}
We see that this is a particular case of the \none equation (\ref{uone})
at $\det h=-1$.  

Moreover, when $\det h$ is not equal to its \ntwo value $(-1)$ Eq.~(\ref{uone})
coincides with Eq.~(\ref{utwo}) after the rescaling
\begin{equation}
u= (-\det h)^{1/3}\, u'\,, \quad m= (-\det h)^{2/3}\, m'\,, 
\quad v= (-\det h)^{-1/3}\,v'\,.
\label{rescale}
\end{equation}
This is in agreement with the master parameter $\sigma$ which contains the
product $m^{-3/2}\det h$ and the nonanomalous U(1) symmetries
we  discussed
above. In other words, breaking of \ntwo by Yukawa couplings does not influence
consideration of the chiral condensates modulus the rescaling~(\ref{rescale}).

The consideration above shows that the only nonperturbative input needed
to determine  the chiral condensates
 is provided by the one-instanton ADS
superpotential. This  means that any reference
 to the \ntwo limit is not crucial at all,
i.e.  in regard to these condensates the exact Seiberg-Witten solution of \ntwo
is equivalent to the ADS superpotential.

The relations for the condensates we have
derived are not new, they were obtained in 
\cite{giveon} by the  `integrating in' procedure introduced in 
\cite{Intrilligator}.
Our approach which is based on `integrating out', plus the Konishi
relations, can  be 
viewed as an independent proof of the `integrating in' procedure.

What we see as an advantage of our approach it is that,
within a   certain range of
parameters, the superpotential (\ref{sup1}) gives a complete
 description of the low
energy physics.  Indeed, when the mass $m_V$ of the field $V$,
\begin{equation}
m_V= 2m\left(2-3\sigma \kappa\right)\;,
\end{equation}
is much less than the other masses, such as $m_\Phi=g^2\mu$ and $m_W=|g^2
v|^{1/2}$, we are in the weakly coupled Higgs phase and enjoy  full
theoretical
control. The Konishi relations help to determine the condensates of heavy
fields in  this
phase.  Holomorphy then allows for continuation of these
results for the condensates to strong coupling.

At strong coupling the superpotential 
(\ref{sup1}), like other versions of the Veneziano-Yankielowicz
Lagrangians~\cite{vy}, 
does not describe the low energy physics. 
For example, it contains no light monopole
degrees of freedom near the monopole vacuum point at small $\mu$.
Moreover, there is no single local superpotential which could
describe mutually nonlocal degrees of freedom which
become light in different regions of the moduli space of
the theory.  At strong coupling the superpotential (\ref{sup1}) is equivalent 
to the effective superpotential~\cite{giveon}
 of the `integrating in' procedure and 
can be viewed as a shorthand equation that gives the values
of the condensates.  

\subsection{Matter and gaugino condensates in the
 limit of large mass}

Here we summarize the results for matter, $v=\langle Q_{ f}^\alpha 
Q^{f}_\alpha/2
\rangle$,  $u=\langle \Tr\,\Phi^2 \rangle$, and gaugino, $s=\langle \Tr
\,\lambda^2/16\pi^2
\rangle$, condensates in the limit  where the parameter
$\sigma$ defined by Eq.~(\ref{sigma}) is small. This can be achieved in
the limit of large
$m$ if the Yukawa coupling is fixed, or by taking  $\det h$
to be small otherwise. In the charge vacuum:
\begin{eqnarray}
&&v_{C}=\frac{2\,\mu\, m}{(-\det h)}\cdot\left(1+{\cal
O}(\sigma^2)\right),\nonumber\\[1mm] 
&& u_C=\frac{m^2}{(-\det h)
}\cdot\left(1+{\cal O}(\sigma^2)\right),\nonumber\\[1mm]
&&s_C=\frac{\mu\,\Lambda_1^3(-\det h)}{8\,m}
\cdot\left(1+{\cal O}(\sigma^2)\right).
\label{charge}
\end{eqnarray}
In the monopole and dyon vacua: 
\begin{eqnarray}
&&v_{\,M,D}=\pm\,\mu\,\sqrt{\frac{\Lambda_1^3}{4m}} 
\cdot\left(1+{\cal O}(\sigma)\right),\nonumber\\[1mm]
&& u_{\,M,D}=\pm \, \sqrt{\Lambda_1^3 m} \cdot\left(1+{\cal
O}(\sigma)\right),\nonumber\\[1mm] 
&&s_{\,M,D}=\pm \, \frac 1 2\,\mu\,
\sqrt{\Lambda_1^3 m} \cdot\left(1+{\cal O}(\sigma)\right).
\label{mondy}
\end{eqnarray}
The upper sign refers to the monopole vacuum, while 
 the lower one is for the dyon
vacuum. As  discussed above we can interpret these vacua also as the 
 two vacua of
the Higgs phase in SQCD.  To this end we need to consider the limit of small 
$\det h$
and $m\ll \Lambda_{\rm SQCD}$ with the identification
\begin{equation}
\Lambda_{\rm SQCD}^5=\frac 1 4 \,\mu^2 \Lambda_1^3
\end{equation}

\subsection{Small mass  limit}

The limit of massless fundamentals $m \to 0$ corresponds to $\sigma \to 
\infty$.  In this limit the 
three vacua are related by a $Z_3$  symmetry~\cite{sw},
\begin{eqnarray}
&& v=\frac{\mu\,\Lambda_1}{(2\det h)^{1/3}}\,e^{2\pi i k/3}
\cdot\left(1+{\cal O}(\sigma^{-2/3})\right), \qquad (k=0,\pm 1),\nonumber\\[1mm]
&& u= \frac 3 8 \, \Lambda_1^2 \left(2\det h\right)^{1/3}\,e^{-2\pi i k/3}
\cdot\left(1+{\cal O}(\sigma^{-2/3})\right),
 \qquad (k=0,\pm 1),\nonumber\\[1mm]
&& s=\frac 1 4 \, \mu\, \Lambda_1^2 \left(2\det h\right)^{1/3}\,e^{-2\pi i
k/3}
\cdot\left(1+{\cal O}(\sigma^{-2/3})\right), \qquad (k=0,\pm 1)
\label{smc}
\end{eqnarray}
Note that the  massless limit exists due to the nonvanishing Yukawa coupling.
When $h\to 0$ we are back to the runaway vacua of massless SQCD.

\subsection{Argyres-Douglas points}

When the mass $m$  changes from large to small values we interpolate 
 between the two
quite different structures of vacua shown above. Let us consider this 
transition
when, for definiteness,
 \mbox{$\det h\!=\!-1$} and $m$ is real and positive
and changes 
  from large to small  values. 
 At large positive $m$ all the vacua are
situated at real values of  $u$, from Eqs.~(\ref{charge},\,\ref{mondy}) we see 
that
the dyon vacuum is at negative $u$, the monopole vacuum is at positive
$u$, and the
charge vacuum is also at positive, but much larger, values of $u$.  When $m$ 
diminishes then  at some point the  monopole and charge vacua collide on the
real axis of
$u$ and subsequently  go more off to  complex values
  producing the $Z_3$ picture at small $m$. 

The point in the parameter manifold where the two vacua coincide is the AD
 point~\cite{ad}. 
In the SU(2) theory these points were studied in \cite{apsw}.
Mutually non-local states, say charges and monopoles, becomes
massless at these points. On the Coulomb branch of the \ntwo theory
these points correspond to a non-trivial conformal field theory
\cite{apsw}. Here we study   the \none SUSY theory, where \ntwo is broken 
by the mass term for the adjoint matter as well as by the difference of the
Yukawa coupling from its \ntwo value.    Collisions of two vacua 
still occur in this theory.  In this subsection we  find the values of $m$
at which AD points appear and 
calculate the values of the condensates at this point. 
 In the next section we study
what happens to the confinement of charges in the monopole point at non-zero
$\mu$ once we approach the AD point.

First, let us work out the AD values  of $m$, generalizing the
consideration in~\cite{apsw}. Coalescence of two roots for $v$  means
that together with Eq.~(\ref{vaceq}) the derivative of its left-hand-side
should also vanish,
\begin{equation}
m-\frac{(-\det h)}{2} \, \frac{v}{\mu}- \frac{\Lambda_1^3}{4} \left(
\frac{\mu}{v}\right)^2=0, \qquad -(-\det h) + \Lambda_1^3 \left(
\frac{\mu}{v}\right)^3=0\;.
\label{vaceq1}
\end{equation}
This system is consistent only at three values of $m=m_{\rm AD}$,
\begin{equation}
m_{\rm AD}=\frac 3 4 \, \omega\, \Lambda_1\,(-\det  h)^{2/3},
\qquad \omega={\rm e}^{2\pi i n/3}\quad (n=0,\pm1)
\;,
\label{mad}
\end{equation}
related by $Z_3$ symmetry. The condensates at the AD vacuum are
\begin{eqnarray}
&&v_{\rm AD}=\omega \,\frac{\mu\,\Lambda_1}{(-\det
h)^{1/3}}\,,\nonumber\\[1mm]
&& u_{\rm AD}= \omega^{-1} \, \frac 3 4  \,  \Lambda_1^2\,  (-\det
h)^{1/3}\,,\nonumber\\[1mm]
&& s_{\rm AD}=\omega^{-1} \, \frac 1 4 \, \mu\Lambda_1^2 \,  (-\det h)^{1/3}\,.
\label{cad}
\end{eqnarray}

\section{Dyon condensates}\label{sec:mcd}

In this section we calculate various dyon condensates at the 
three vacua of the 
theory.
As  discussed above, holomorphy allows us to find these condensates
starting from  a consideration on the Coulomb branch in \ntwo near 
the singularities associated with a given massless dyon. Namely, 
we calculate the monopole
condensate
near the monopole point, the charge condensate near the charge point and the 
dyon
$(n_{m},n_{e})=(1,1)$ condensate near the point where this dyon is light. 
Although
we start with small values of the 
adjoint mass parameter $\mu$,  our results for
condensates are exact for any $\mu$ as well as for any value of $\det h$.

\subsection{Monopole condensate.}

Let us start with calculation of the monopole condensate near the
monopole point. Near this point the effective low energy
description of our theory can be given in terms of \ntwo dual QED~\cite{sw}. It
includes a light monopole hypermultiplet interacting with a 
vector (dual) photon
multiplet in the same way as electric charges interact with ordinary photons.
Following Seiberg and Witten~\cite{sw} we write down the effective 
superpotential
in the following form,
\beq
W= \sqrt{2}\,\tilde{M}MA_{D}+\mu\, U,
\label{mqed}
\eeq
where $A_{D}$ is a neutral chiral  field (it is a part of the \ntwo
dual photon multiplet in the \ntwo theory) and $U\!\!=\!\Tr \Phi^2$
considered as a function of $A_{D}$. The
second  term 
breaks \ntwo supersymmetry down to \none.

Varying  this superpotential with respect to $A_{D}$, $M$
and $\tilde{M}$ we find that $A_{D}=0$, i.e.  the monopole mass vanishes, and
\beq
\langle\tilde{M}M \rangle=-\left.\frac{\mu}{\sqrt{2}}\frac{{\rm d}{u}\:}{{\rm
d}{a_{D}}}\right|_{a_{D}=0}\,.
\label{mc}
\eeq
The condition $A_{D}=0$ means that the Coulomb branch near the
monopole point, where the monopole mass vanishes, shrinks to a single vacuum
state at the singularity while Eq.~(\ref{mc}) 
determines the value of monopole condensate. Below we consider $a_D$
as a function of $u$. The value of $u$, $u=u_M$, at the monopole
vacuum was determined in the
previous section.

The non-zero value of the monopole condensate ensures 
U(1) confinement for charges via the formation of
Abrikosov-Nielsen-Olesen  vortices. Let us work out
the r.h.s. of Eq.~(\ref{mc}) to determine the $\mu$
and $m$ dependence of the monopole condensate.
From the exact Seiberg-Witten solution \cite{sw}, we have
\beq
\frac{{\rm d}{a_{D}}}{{\rm d}u}=
\frac{\sqrt{2}}{8\pi }\oint_\gamma \frac{{\rm d}x}{y(x)}\,.
\label{con}
\eeq
Here for $y(x)$  given by Eq.~(\ref{swu}) we use the form
\beq
y^2=(x-e_{0})(x-e_{-})(x-e_{+})\,.
\eeq
The integration contour $\gamma$ in the $x$
plane  circles around  two branch points  $e_{+}$
and $e_{-}$ of $y(x)$. At the monopole vacuum, when  $u=u_{M}$, two
branch points 
$e_{+}$ and $e_{-}$ coincide, $e_{+}=e_{-}=e$ and the integral~(\ref{con}) is 
given
by the residue at $x=e$. 
\beq
\frac{{\rm d}{a_{D}}}{{\rm d}u}(u_{M}) =
\frac{i\,\sqrt{2}}{4\,\sqrt{e-e_0}}\,.
\eeq
The value of $e-e_0$ (equal at $u\!=\!u_M$ to $(1/2)\,{\rm d}^2 (y^2)/{\rm 
d}x^2$ )
is fixed by the equation ${\rm d}(y^2)/{\rm d}x=0$,
\beq
e-e_{0}=\sqrt{u_{M}^2-\frac 3 4 m\Lambda_{1}^{3}}\;.
\eeq
Substituting this into the expression for the monopole condensate
(\ref{mc})  we get finally
\beq
\langle \tilde M M\rangle =2i\mu\left(u_{M}^2-\frac 3 4
m\Lambda_{1}^{3}\right)^{1/4}.
\label{mm}
\eeq

To test the result let us consider first the limit of a 
large masses $m$ for the fundamental matter. 
 As  in Sec.~\ref{sec:sup} this limit can be viewed as a
RG flow to pure Yang-Mills theory with the identification
\beq
\Lambda_{0}^{4}=m\Lambda_{1}^3,
\eeq
where $\Lambda_{0}$ is the scale of the \ntwo Yang-Mills theory.
In this limit we have $u_{M}=\Lambda_{0}^2$. Then Eq.~(\ref{mm})
gives
\beq
\langle \tilde M M\rangle =\sqrt{2}\,i\,\mu\,\Lambda_{0}\,,
\eeq
which coincides with the Seiberg-Witten result~\cite{sw}. This ensures monopole
condensation and charge confinement in the monopole point at large $m$.

Notice, that in the derivation above \ntwo was not broken by the Yukawa 
coupling,
i.e. we assume $\det h=-1$. 
The result, however, can be easily generalized
to arbitrary 
$\det h$ by means  of U(1) symmetries considered above. 
The U$_{R}(1)$ charge of $\tilde M M$ is equal to one.
Indeed, the coefficient of the $\sqrt{2}\tilde M M A_D$ term in the
superpotential~(\ref{mqed}) which is equal to one in the \ntwo limit
remains the same when \ntwo is broken down to \none. It follows from 
U(1) symmetries together with decoupling of fundamental matter at
large $m$. As a result 
 we see that the 
Eq.~(\ref{mm}) for the  monopole condensate remains
valid for arbitrary $\det h$.

It is instructive to rewrite the result~(\ref{mm}) for the monopole
condensate in terms of $v$,
\begin{equation}
  \label{mmV}
  \langle\tilde M M \rangle=i\left[ (-\det h)\,
 v^2- \frac{\mu^3
      \Lambda_1^3}{v}\right]^{1/2}
=i
\left[(-\det h)\, v^2- 4\mu \, s\right]^{1/2}\,,
\end{equation}
where we also show a form which uses the  gluino condensate $s$. It
follows from  the expression~(\ref{ukappa}) for  $u$ and
Eq.~(\ref{vaceq}) for $v$.  It is
interesting to observe that at $m\ll \Lambda_1$, when the value of $v$ is large
and the nonperturbative
term in (\ref{mmV}) can be neglected, the monopole
 condensate reduces to that of the  quark,
 $\langle\tilde M M \rangle \to iv\sqrt{-\det h}$. 
Another interesting limit is the SQCD one when $h\to 0$. In this limit
the nonperturbative term in Eq.~(\ref{mmV}) dominates, and
the square of the monopole condensate reduces to the gluino
condensate.

Now let us address the question: what happens with the monopole
condensate when we reduce $m$ and approach the AD point?
The AD point corresponds to a particular
value of $m$ which ensures  coalescence of the monopole and charge
singularities in the $u$ plane. Near the monopole point we have
condensation of monopoles and confinement of charges while
near the charge point we have condensation of charges and
confinement of monopoles. As  shown by 't~Hooft 
these two phenomena cannot happen simultaneously~\cite{H}. The
question is:  what happens when monopole and charge points collide
in the $u$ plane?

The monopole condensate
 at the AD point is given by Eq.~(\ref{mm}). When $m$ and $u$ are
 substituted by 
 $m_{AD}$ and $u_{AD}$ from Eqs.~(\ref{mad}) and (\ref{cad}), we get
\beq
\langle \tilde M M\rangle_{AD}=0.
\eeq
We see that the monopole condensate goes to zero 
at the AD point. Our derivation makes it clear why it happens.
At the AD point all three roots of $y^2$
become degenerate, $e_+=e_-=e_0$, so the monopole condensate which is
proportional to $\sqrt{e-e_0}$ naturally vanishes.  

In the next subsection we
calculate the charge condensate in the charge point and show that it  also 
goes to
zero as $m$ approaches its AD value~(\ref{mad}). Thus we interpret
the AD point as a deconfinement point for both monopoles
and charges.

\subsection{Charge and dyon condensates}

In this subsection we use the same method to calculate
values for the  charge and dyon condensates near the charge and dyon points
respectively. We first consider $m$ above its AD value (\ref{mad}) and
then continue our results to values of $m$ below $m_{AD}$. In
particular, in the limit $m=0$ we recover $Z_{3}$ symmetry.

Let us start with the charge condensate. At $\mu=0$, $\det h=-1$ and large $m$ 
the
effective theory near the charge point 
\begin{equation}
a=-\sqrt{2}\,m
\label{chp}
\end{equation}
on the Coulomb branch
is \ntwo QED. Here $a$ is the neutral scalar, the partner of photon
in the \ntwo supermultiplet.
 Half the degrees
of freedom in  color doublets  become massless whereas the other
half  acquire large a mass $2m$. The massless fields form one hypermultiplet
$\tilde Q_+, Q_+$ of charged particles in the effective electrodynamics.
 Once we add the mass term for the adjoint matter
 the effective superpotential near the charge point becomes
\beq
{\cal{W}}=\frac{1}{\sqrt{2}}\,\tilde Q_{+}Q_{+}A+m\,\tilde Q_{+}Q_{+}
 +\mu\, U 
\eeq
Minimizing this superpotential we get condition (\ref{chp})
as well as
\beq
\langle\tilde{Q}_{+}Q_{+}\rangle= -\left. \sqrt{2}\,\mu\,\frac{{\rm d} u}{{\rm d} a}\right|_{a=-\sqrt{2}\,m}\,.
\eeq
Now, following the same steps which led us from (\ref{mc}) to
 (\ref{mm}), we get
\beq
{\sqrt{-\det h}}\, \langle\tilde Q_{+}Q_{+}\rangle=2\,\mu\,
(u_{C}^2-\frac 3 4 
m\,\Lambda_{1}^{3})^{1/4}\,,
\label{chc}
\eeq
where we include a generalization to arbitrary $\det h$.
We choose to consider\\ ${\sqrt{-\det h}}\, \langle\tilde
Q_{+}Q_{+}\rangle$ because it has the U$_R$(1) charge equal to one,
similar to the  $\langle\tilde M M \rangle$ condensate considered above.
By $u_{C}$ we denote the position of the charge vacuum 
in the $u$ plane. At large $m$ $u_{C}=m^2/(-\det h)$, 
 see Eq.~(\ref{charge}), and
\beq
\langle\tilde Q_{+}Q_{+}\rangle =\frac{2\,\mu \,m}{(-\det h)}
\left(1+{\cal O}(\sigma^2)\right)\,.
\label{chcl}
\eeq

Holomorphy allows us to extend the result (\ref{chc}) to arbitrary $m$ and
$\det h$. So we can use Eq.~(\ref{chc}) to find  the charge condensate at 
the AD
point. Using Eqs.~(\ref{mad}) and (\ref{cad}) we see that the 
charge condensate
vanishes at the AD point in the same manner the monopole condensate
does. As it was 
mentioned we interpret this as deconfinement for both charges and monopoles.

As with the monopole condensate, we can also relate the
charge condensate with the quark vev $v$,
\begin{equation}
  \label{ccV}
  \langle\tilde Q_+ Q_+ \rangle= \left[v^2- \frac{\mu^3
      \Lambda_1^3}{v(-\det h)}\right]^{1/2}
=\left[ v^2- \frac{4\mu \, s\,}{(-\det h)}\right]^{1/2}\,,
\end{equation}
This expression differs from the one for the monopole condensate only
by a phase factor.
The coincidence of the charge condensate with the quark one at large
$v$, i.e. at weak coupling, is natural. The difference is due to
nonperturbative effects and is  similar to the difference between
$a^2/2$ and  $u$ on
the Coulomb branch of the \ntwo theory. At strong coupling the
difference is not small. In particular, the charge condensate vanishes
at the AD point while the quark condensate remains finite. 

Now let us work out the dyon condensate. More generally let us
introduce the dyon field $D_{i}\,$, $i=1,2,3$, which
stands for charge, monopole and $(1,1)$ dyon,
\beq
D_{i}=\left\{(-\det h)^{1/4} Q_{+},\; M,\; D\right\}.
\eeq
The arguments of the previous subsection
which led us to the result (\ref{mm}) for monopole condensate
gives for $\langle\tilde D_i \, D_i\rangle$ 
 \beq
\langle\tilde D_i \, D_i
\rangle=2\,i\,\zeta_{i}\,\mu
\left(u_{i}^2-\frac 3 4
\,m\,\Lambda_{1}^{3}\right)^{1/4},
\label{dyc}
\eeq
where $u_{i}$ is the position of the i-th point in the
$u$ plane and the  $\zeta_{i}$ are  phase factors. 

For the monopole condensate  at real values of $m$ larger than
the $m_{\rm AD}$  Eq.~(\ref{mm}) gives
\beq
\zeta_{M}=1,
\eeq
while for the charge condensate from Eq.~(\ref{chc}) we have
\beq
\zeta_{C}=-i.
\eeq
In fact one can fix the charge phase factor by  imposing
the condition that the charge condensate should approach
the value $2m \mu$ in the large $m$ limit.
For the dyon the phase factor is 
\beq
\zeta_{D}= i\,.
\label{dph}
\eeq

At the particular AD point we have chosen the monopole and
charge  condensates vanish, while
the dyon condensate remains non-zero, see (\ref{dyc}). Below the AD point,
condensates are still given by  Eq.~(\ref{dyc}), but the
charge and monopole phase
factors  can change~\footnote{Note
that the quantum numbers of the ``charge'' and ``monopole'' are also
transformed, see \cite{BF}}. The dyon phase factor~(\ref{dph})
does not change when we move through the AD point  because the dyon
condensate does not vanish at this point.

In the limit $m=0$ we should recover the $Z_3$-symmetry for the
values of condensates. From Eq.~(\ref{dyc}) it is clear that the
absolute values of all three condensates are equal because
the values of the three roots $u_{i}$ are on the circle in the
$u$ plane, see (\ref{smc}). Imposing the requirement of
$Z_3$ symmetry at $m=0$ we can fix the unknown phase factors
$\zeta_{C}$  and $\zeta_{M}$ below the AD point using the value
 (\ref{dph}) for dyon.  This gives
\beq
\zeta_{C}= i\,  ,\quad
\zeta_{M}=- i\,.
\eeq

\subsection{Photino and gaugino condensates}

The gaugino condensate $\langle\Tr \lambda^2\rangle$ we found in the
previous section can be viewed as a sum of the 
condensates for charged gauginos and the 
photino,
\begin{equation}
\left\langle\Tr \lambda^2\right\rangle=
\left\langle\lambda^+\lambda^-\right\rangle +
\frac 1 2\,
\left\langle\lambda^3\lambda^3\right\rangle
\end{equation}
In  gauge invariant form the photino condensate can be associated with 
\begin{equation}
\left\langle(\Tr W \Phi)^2\right\rangle
\end{equation}
We argue here that the photino condensate vanishes so that the gaugino
condensate is solely due to the charged gluino.

Let us start with the Coulomb branch in the 
\ntwo theory.  All gaugino condensates 
 vanish in \ntwo for a simple reason: $\lambda^2$ is {\em not} the lowest
component in the corresponding \ntwo supermultiplet. When the
perturbation   $\mu\, U$ which breaks \ntwo is added to the superpotential
the  gaugino condensate is
proportional to $\mu$. However, the term $\mu \, U$ in the superpotential 
does not break \ntwo SUSY in the effective QED. Consider, for example, the
monopole vacuum. The corresponding effective superpotential is given by
Eq.~(\ref{mqed}), where in the expansion of $U$ as function of $A_D$ it is 
sufficient
to retain only linear term.  It was shown in~\cite{HSZ} 
 that the perturbation linear in $A_D$   does not break \ntwo in the effective 
QED.
An immediate consequence of this observation is that the photino
condensate continues to  vanish. 

\section{The Argyres-Douglas point: how well is the theory  defined}
\label{sec:walls}

As  discussed in the Introduction, at the AD point we encounter the
problem of not having a uniquely defined vacuum state. Indeed, when the mass
parameter $m$ approaches its AD value $m_{AD}$ we deal with two vacuum
states which can be distinguished by values of the chiral condensates. It
is unlikely that the number of states with zero energy will  change
when we reach the AD point, it is very  similar to the Witten index.
However, the continuity of the chiral condensates we obtained above shows 
that they are no longer parameters which differentiate the two states once
we reach the AD point.

This does not prove the  absence of a relevant order parameter so the quest
can be continued. A natural possibility to consider is a domain walls
interpolating between colliding vacua. In the case of BPS domain walls
their tension is given by the central charge~\cite{ds},
\begin{equation}
T_{ab}=2\,|{\cal W}_{\rm eff}(v_a) - {\cal W_{\rm eff}}(v_b)|
\end{equation}
where $a$,$b$ label the colliding vacua. The central charge here is
expressed via values of exact superpotential~(\ref{sup1}) in
corresponding vacua. The continuity of the condensate $v$ shows that 
the domain wall becomes tensionless at the AD point, 
$T\propto (m-m_{\rm AD})^{3/2}$ when $m\to m_{\rm AD}$.  If such a domain
wall were observable at the AD point it could serve as a signal of two vacua.

We argue, however, that this domain wall is not observable in continuum
limit. The crucial point is that the wall is built  out of massless
fields, therefore its thickness is infinite at the AD point. This makes
it impossible to observe this tensionless wall in any physical
experiment of a limited spatial scale.

In the conclusion of this section let us review briefly the brane
construction of  \none vacua.  Gauge theories are realized on
 brane worldvolumes. Brane configurations responsible for  \none
 theories were suggested in~\cite{N_one} and a derivation of domain wall
 tensions from analysis of Riemann surfaces
(which is similar to the calculation of the masses of BPS particles in \ntwo
theories) can be found in~\cite{super}. The brane configuration for the
\none theory with one flavor and SU(2) gauge group is described by
 Riemann surface embedded into
three dimensional complex space $C^3$ parametrized by three variables
$t$, $v$ and $w$. The embedding is given by the following
equations 
\begin{eqnarray}
  \label{brane}
&&  v+m=\frac{(w-w_{+})(w-w_{-})}{\mu w}\;,\qquad
t=\mu^{-2}w(w-w_{+})(w-w_{-})\;;\nonumber\\[1mm]
&& w_{+}+\frac{1}{2}\,w_{-}+ m \mu=0\;,\qquad\qquad  w_{-}^{2}w_{+}=-(\mu \Lambda_1)^{3}\;.
\end{eqnarray}
with free parameters  $\mu,m,\Lambda$. The tension of the walls, which have
the interpretation of the M5 branes wrapping 
 three-cycle with the boundaries on the Riemann surface above 
can be calculated by
integrating the holomorphic  three-form over this cycle 
\begin{equation}
T=\int{\rm d}v \wedge {\rm d}w \wedge {\rm d}(\log t)
\end{equation}

Let us consider the geometry of the brane configuration near the AD point. It
was shown recently~\cite{svafa} that the AD point corresponds to a
singular Calabi-Yau 3-manifold which is resolved if one adds 
particular perturbation. Since the tension is defined by integration
of the holomorphic 3-form  around the resolved singularity  the
 tensionless wall has  a geometrical interpretation as the M5 
brane wrapping this vanishing cycle. Actually, the curves can be considered as
fibered  over
the complex $m$ plane and the AD singularities correspond to the
appearance of  vanishing cycles in the fiber in a manner quite similar to the
 Seiberg-Witten solution of \ntwo
theories where  vanishing cycles correspond to  massless BPS particles.

\section{Conclusions}\label{sec:disc}

The approach of this work is similar to that used in SQCD. Namely, we 
integrate
out the adjoint field which leads, in some range of parameters, to
an  SQCD-like
effective superpotential. This superpotential describes
 the low energy theory at
weak coupling where we have  full theoretical control.  The nonperturbative 
part
 is given by the ADS superpotential generated at the  one instanton level.
 The adjoint field 
shows up only as an extra (as compared with SQCD) nonrenormalizable term
quartic in the fundamental fields. 

Results for chiral condensates of  matter and gaugino 
fields   are
continued 
into the range of a small adjoint mass where we find a complete matching with 
the \ntwo Seiberg-Witten solution. The Argyres-Douglas points
introduced in \ntwo theories are shown to exists in the \none theory as
well. 
Although the bulk of our results for matter 
and
gaugino condensates overlaps with what is 
known in the literature we think that our
approach clarifies some aspects of duality in \none theories.

We then analyze  monopole, charge and dyon condensates departing from 
the Coulomb branch of the \ntwo theory. This resulted
 in  explicit relations 
between
these condensates and those of the fundamental matter. The most
interesting 
phenomenon occurs at the AD point: when the monopole and charge
vacua collide both the monopole and charge condensates vanish. We interpret 
this as a deconfinement of electric and magnetic charges at the AD  
point. Vanishing of condensates signals that the theory at this point
becomes superconformal.

In our approach we see straightforwardly that the one-instanton
generated ADS superpotential is the only nonperturbative input needed
to fix all chiral condensates. The general nature of this statement 
is seen from our derivation which relates polynomial coefficients in
the  Seiberg-Witten curve to the ADS superpotential.
 
Let us mention a relation to  finite-dimensional
integrable systems.
It was recognized that \ntwo theories are governed by  finite-dimensional
integrable systems. The integrable system responsible for \ntwo SQCD
was identified with the nonhomogenious XXX spin chain~\cite{gm}.
After perturbation to the \none theory the Hamiltonian of the integrable
system is expected to coincide with the superpotential of 
corresponding \none theory. This has been confirmed by direct calculation 
in the pure \ntwo gauge theory~\cite{vafa} as well in the  theory with
a massive adjoint multiplet~\cite{dorey}. It would be very 
interesting
to find a similar connection between spin chain Hamiltonians and
superpotentials in the \none SQCD. One more point to be clarified is the
meaning of the AD
point within approach based on integrability. Since the quark mass is 
identified as a value of spin~\cite{gm} one might expect that at
particular spin values  corresponding to the AD mass, the  XXX spin chain
would have  additional symmetries similar to superconformal ones. We hope
to discuss these points in more details elsewhere.

In this paper we considered only the SU(2) theory with one flavor
postponing the generic $N_c$, $N_f$ case for a separate publication.
The most interesting problem in the generic situation involves Seiberg
IR duality of the electric  SU$(N_c)$ theory with $N_f$ flavors and
the magnetic SU$(N_f-N_c)$ theory.
In generic case of nondegenerate fundamental masses  we  expect deconfinement
at the AD points. A degeneracy in fundamental masses leads to the 
  appearance of Higgs branches. The approach of the present paper
 can be applied to this case as well.
 However,
since Higgs branches do not disappear at the AD points~\cite{apsw} we
do not expect deconfinement to occur in this case~\cite{Y}.

\subsection*{Acknowledgments}
Authors are grateful to P.~Argyres, A.~Hanany, K.~Konishi, A.~Marshakov,
 S.~Rudaz, A.~Ritz, 
and
M.~Shifman for helpful discussions. Part of this  work was done when two of the
authors, A.V. and A.Y., participated in the SUSY99 program organized by
the Institute for Theoretical Physics at Santa Barbara. A.V. and A.Y. are 
thankful to ITP for  hospitality and support from NSF by the grant
PHY 94-07194. A.G. and A.Y. thank the Theoretical Physics Institute 
at the University of Minnesota where this work was initiated for
  support. A.G. thanks
J. Ambjorn for  hospitality at the Niels Bohr Institute where 
part of  this work has been done.

The work of A.G. is  supported in part by the grant INTAS-97-0103,
A.~V. is  supported in part by DOE under the grant DE-FG02-94ER40823, 
and A.Y. is supported in part by Russian Foundation for
Basic Research under the grant  99-02-16576.

\end{document}